# Comments on "All-electron self-consistent *GW* in the Matsubara-time domain: Implementation and benchmarks of semiconductors and insulators"

Diola Bagayoko,[1] Yacouba Issa Diakité,[2] Lashounda Franklin[1]

[1]Department of Mathematics and Physics, Southern University and A&M College
in Baton Rouge (SUBR), Baton Rouge, LA 70813, USA
[2]Faculté des Sciences et Techniques (FST), Université des Sciences,
des Techniques, et des Technologies de Bamako (USTTB), Bamako, Mali

Chu *et al.* recently reported extensive results of local density approximation (LDA) and of four (4) different Green function and dressed Coulomb (GW) approximation calculations of electronic properties of several semiconductors and insulators [Phys. Rev. B 93, 125210 (2016)]. Due to their oversight of more than a dozen previous, ab-initio LDA findings, their LDA results constitute an unintended misrepresentation of the capabilities of LDA in correctly describing electronic and related properties of materials. These comments are to address this oversight and to show, contrary to the findings and assertions of Chu *et al*, that LDA calculations do predict and describe accurately electronic and related properties of semiconductors – provided these calculations adhere to conditions of validity of DFT. Additionally, the referenced, previous LDA results are mostly in excellent agreement with experiment, unlike most of their corresponding ones from the fully self-consistent GW results of Chu *et al.*

Chu *et al.*[1] recently performed extensive calculations of electronic properties of several semiconductors. They employed a local density approximation (LDA) potential, for the density functional theory (DFT) calculations, and four implementations of the Green function and dressed Coulomb approximation (GWA). Specifically, in addition to the single shot, non self-consistent $G_0W_0$ calculations, they performed partially self-consistent ($GW_0$) and fully self-consistent (sc-GW) computations for the same semiconductors and insulators. The $G_0W_0$, $GW_0$, and sc-GW calculations were performed within the Matsubara-time domain; further, these calculations were carried out using a recent method of analytic continuation, i.e., the continuous-pole expansion (CPE), and the Padé approximation. Even though the two approaches are found to lead to very similar results, in general, the continuous-pole expansion is reported to provide a better description of electronic properties of strongly correlated systems. Chu *et al.*[1] noted the limitations of $G_0W_0$, including violations of momentum, energy, and particle conservation laws. While the calculations were carried out using an all-electron computational package, the $G_0W_0$ calculations were also done in the pseudo potential approximation (PPA) framework.

Table I below shows the results of the extensive calculations for the band gaps of 18 semiconductors and insulators. The first seven (7) columns of the table are from Table IV of Chu *et al.*, including References 2-12 for experimental works. Column eight (8) contains density functional theory (DFT) results, from References 13-23, obtained mostly with the local density approximation potential of Ceperley and Alder[24] as parameterized by Vosko *et al.*[25] The experimental results, except for the underlined ones, are as given by Chu *et al.*[1] It is apparent in Table I that the LDA results of Chu *et al.* (in Column 2) generally underestimate the measured band gaps (in Column 7). While stating this fact, as done by Chu *et al.*, is expected, the oversight of the 14 DFT results in Column 8 leads to an inaccurate view of the capabilities of DFT, inasmuch as the latter calculated results are mostly in agreement with corresponding, experimental ones.

**TABLE I**. Calculated, electronic band gaps (in eV) of selected semiconductors and insulators. The Columns of results are labeled with the potential and computational features as shown below. The results for the DFT-LDA and different levels of Matsubara-time $GW$ ($G_0W_0$, $GW_0$, and full-$GW$), and PPA-$G_0W_0$ are from Chu *et al*, with values between parentheses obtained with the Padé approximation. The experimental results, except for the underscored ones, are from Table IV of Chu *et al*.

|  | DFT-LDA | $G_0W_0$ | $GW_0$ | Full sc-$GW$ | PPA-$G_0W_0$ | Expt. | LDA-BZW-EF |
|---|---|---|---|---|---|---|---|
| Si | 0.58 | 1.38 (1.36) | 1.59 (1.58) | 1.44 (1.44) | 1.28 | 1.25[a] | 1.02[l] |
| Ge | 0.03 | 0.58 (0.59) | 0.85 (0.86) | 0.85 (0.85) | 0.71 | 0.74[a] | 0.644[m] (0.65 GGA) |
| GaAs | 0.24 | 1.48 (1.47) | 1.82 (1.83) | 1.80 (1.81) | 1.51 | 1.52[b] | 1.429[n] |
| SiC | 1.27 | 2.44 (2.45) | 2.90 (2.90) | 2.64 (2.56) | 2.30 | 2.40[a] | 2.24[o] |
| CaSe | 2.00 | 3.89 (3.94) | 4.60 (4.64) | 4.35 (4.34) | 3.89 | 3.85[c] |  |
| C | 4.14 | 6.15 (6.15) | 6.42 (6.43) | 6.10 (6.11) | 6.09 | 5.48[a] | 5.05[l] |
| NaCl | 4.74 | 8.09 (8.11) | 9.00 (9.02) | 8.27 (8.28) | 8.11 | 8.5[d] |  |
| MgO | 4.65 | 7.79 (7.78) | 8.74 (8.74) | 7.94 (7.94) | 7.75 | 7.83[e] |  |
| BN | 4.34 | 6.71 (6.73) | 7.16 (7.18) | 7.10 (7.11) | 6.58 | 6.1–6.4[f] | 6.48[p] |
| LiF | 8.94 | 14.51(14.54) | 15.78 (15.81) | 14.45 (14.47) | 14.55 | 14.20[g] |  |
| SrTiO3 | 1.75 | 3.58 (4.08) | 7.01 (7.13) | 6.87 (7.22) | 3.86 | 3.25[h] | 3.24[q] (GGA) |
| Cu2O | 0.52 | 1.61 (1.54) | 2.16 (2.17) | 2.00 (2.02) | 1.59 | 2.17[i] |  |
| GaN | 1.70 | 3.01 (3.05) | 3.61 (3.66) | 3.36 (3.38) | 3.03 | 3.27[j] | 3.2[l] 3.289[r] |
| ZnO | 0.60 | 2.31 (2.35) | 3.69 (3.71) | 3.53 (3.56) | 2.32 | 3.44[k] | 3.39[s] |
| ZnS | 1.80 | 3.46 (3.43) | 4.06 (4.09) | 3.92 (3.85) | 3.43 | 3.91[k] | 3.725[t] |
| ZnSe | 1.01 | 2.43 (2.48) | 3.03 (3.09) | 2.94 (2.96) | 2.50 | 2.95[a] | 2.6[u] |
| CdS | 0.86 | 2.01 (2.03) | 2.63 (2.66) | 2.49 (2.50) | 2.06 | 2.50[a] | 2.39[v] |
| CdSe | 0.34 | 1.42 (1.51) | 1.97 (1.98) | 1.92 (1.93) | 1.46 | 1.83[a] |  |

[a]Reference [2]
[b]Reference [3]
[c]Reference [4]
[d]Reference [5]
[e]Reference [6]
[f]Reference [7]
[g]Reference [8]
[h]Reference [9]
[i]Reference [10]
[j]Reference [11]
[k]Reference [12]
[l]Reference [13]
[m]Reference [14]
[n]Reference [15]
[o]Reference [16]
[p]Reference [17]
[q]Reference [18]
[r]Reference [19]
[s]Reference [20]
[t]Reference [21]
[u]Reference [22]
[v]Reference [23]

According to Bagayoko,[26] the differences between the two sets of DFT results are not due to any specific deficiency of DFT, rather, they stem from the degree to which the respective calculations adhered to well defined conditions for their results to possess the full, physical content of DFT. While it is presumed that all the calculations met the first of these conditions, i.e., the conservation of the total number of particles, there is no indication in their article that the LDA calculations of Chu et al adhered to the other conditions. The first of these other conditions is (a) that either the calculations utilize the ground state charge density of the system, *à priori* unknown, or (b) that they search and verifiably attain the absolute minima of the occupied energies of the systems under study. The last one is that the calculation whose results possess the full, physical content of DFT is the one with the smallest basis set among the potential infinite ones that lead to the absolute minima of the occupied energies. This latter requirement stems from a corollary of the first Hohenberg-Kohn theorem, i.e., that the energy content of the Hamiltonian (and equivalently the spectrum of the Hamiltonian) is a unique functional of the ground state charge density.

Fifty years of misunderstanding of a key feature of DFT has resulted in the fact that the last two conditions above have been ignored for the most part. This misunderstanding can be summarized as follows: The necessary minimization of the energy, as required by the second Hohenberg-Khon theorem, *is not attained* by reaching self-consistency following the selection of a basis set, irrespective of how judicious this choice may be. Such a minimization produces a stationary solution. Upon starting with a relatively small basis set, it is necessary to augment the basis set methodically and to perform several, successive calculations. As the basis set increases, both occupied and unoccupied energies are generally lowered. Upon the attainment of the absolute minima of the *occupied energies*, as thoroughly explained elsewhere,[14,20,26] they are no longer lowered by an increase of the size of the basis set. However, due to the Rayleigh theorem, a mathematical artifact results in the continued lowering of some unoccupied energies. As per the first Hohenberg-Kohn theorem, such lowered, unoccupied energies, while the charge density, the Hamiltonian, and the occupied energies no longer change, do not belong to the DFT spectrum of the Hamiltonian which, as per the first theorem, is a unique functional of the ground state charge density.

The above points address the unintended misrepresentation of the capabilities of DFT, due in part to the oversight of results in agreement with experiment. Chu *et al.* invoked the derivative discontinuity[27-29] of the exchange correlation energy as an explanation of the underestimation of measured band gaps by DFT, particularly LDA calculations. The LDA and GGA results in Column 8 suffice to question that explanation, given that the related calculations did not attempt to correct neither for the derivative discontinuity, nor for the effects of self-interaction.

In addition to the numerical results to the contrary of the need for a derivative discontinuity, Bagayoko[26] underscored the mathematical and physical reason that this discontinuity should not have much bearing, if any, on ground state properties. Indeed, the very derivation of the Kohn-Sham equation requires a constant number of particles: i.e., $\delta N = 0$, where N is the total number of particles, is what led Kohn and Sham to their equation. All introductions[27-29] of the derivative discontinuity known to us consider a system where the total number of particle changes by at least one. An entity derived under this condition may have applications for excited state studies, but should not have much bearing on ground state properties. Perdew et al.[27] made it clear that the band gap is a ground state property, given that it is a difference between ground state energies. One should therefore expect to obtain it accurately without invoking the derivative discontinuity. Incidentally, a careful reading of the References 27-29 reveals that while the discontinuity has been established, in the condition noted above, it has not been proven that it is non-zero in real semiconductors or insulators. Perdew and Levy[27] asserted, without a proof, that it is non-zero in real semiconductors; Sham and Schlüter[28-29] explicitly stated that their work does not prove or disprove that the discontinuity is non-zero in real insulators.

The GW approximation has led to significant improvements, over many previous LDA results mostly obtained with a single basis set. This fact is evident in the drastic differences between the LDA results of Chu *et al.*, in Column 2, and their GW ones in Columns 3-6, with the latter generally closer to corresponding, experimental ones. A comparison of these results with the LDA BZW or BZW-EF results, in Column 8 show that the latter are generally closer to corresponding, experimental ones, except for ZnSe others than ours. Specifically, for the first seven (7) materials we studied, from Si to $SrTiO_3$, the LDA BZW or BZW-EF results are visibly closer to experimental ones as compared to those of fully self-consistent GW findings. The difference is particularly large for $SrTiO_3$ for which we obtained 3.24 eV for the experimental value of 3.25 eV, while the sc-GW result is 6.87 eV or 7.22 eV, where the latter is obtained with the Padé approximation. Clearly, LDA calculations that strictly adhere to the necessary conditions for their results to possess the full, physical content of DFT can accurately describe and predict electronic and related properties of materials. As such, such calculations can inform and guide the design and fabrication of semiconductor-based devices, as for the Materials Genome Inititative.[30]